\documentclass{nrlreport} 

\defaultbibliography{2D_LDV_Report.bib}

\nrlheadertitle{Remotely Operating a Single-Point LDV System to Acquire 2D Measurements}
\nrlheaderauthor{Alec Ikei, Dr. Kaushik Sampath}
\nrlbranch{Acoustic Signal Processing and Systems Branch\\Acoustics Division}

\nrlauthortwo{}   \nrlbranchtwo{}

\nrlapprovaldate{January 27, 2021}

\nrlwarning{}  \nrldistribution{Distribution A: Approved for public release: distribution unlimited}
\nrlclassifiedby{}  \nrldeclassifyon{}  \nrldissemination{}

\nrlreporttype{NRL Memorandum Report}
\nrldatescovered{01/01/2019 - 01/07/2021}  \nrlauthorlist{Alec K. Ikei, Dr. Kaushik Sampath}

\nrlcontractnumber{}  \nrlgrantnumber{}  \nrlprogramelement{}
\nrlprojectnumber{}  \nrltasknumber{}  \nrlworkunitnumber{}

\nrlorganizationaddress{U.S Naval Research Laboratory\\4555 Overlook Ave SW\\Washington, DC 20375}  \nrlsponsoraddress{}
\nrlsponsoracronym{}  \nrlsponsorreportnumber{}
\nrlsupplementarynotes{}  \nrlsubjectterms{}

\nrlreportclass{U}  \nrlabstractclass{U}  \nrlthispageclass{U}
\nrlabstractlimitation{UU}  \nrlresponsibleperson{Alec K. Ikei}  \nrltelephonenumber{(202) - 404 - 4816}

\begin{document}
\title{Remote Operation of a Single-Point LDV System to Acquire 2D Measurements}

\author{Alec K. Ikei, Dr. Kaushik Sampath}

\date{\today} 
\maketitle    

\nrlabstract{Due to the unprecedented increase in telework requirements, the motivation to further automate and remotely control experiments has become apparent.  This work documents the technical development of creating a two-dimensional (2D) Laser Doppler Vibrometry (LDV) measurement using a single-point LDV system through an automated and remotely controllable process.  This report aims to assist in rapid development of setups for similar use cases.  The setup described is also modular, and has been used to analyze the modal response of samples actuated through air-based acoustic signals as well as those mechanically induced.} 

\tableofcontents  

\begin{executivesummary}
Due to the unprecedented increase in telework requirements, the motivation to further automate and remotely control experiments has become apparent.  This work documents the technical development of creating a two-dimensional (2D) Laser Doppler Vibrometry (LDV) measurement using a single-point LDV system through an automated and remotely controllable process.  This report aims to assist in rapid development of setups for similar use cases.  A key achievement of this work is the complete control integration of various types of hardware and communication protocols, i.e. Field-Programmable Gate Array (FPGA), function generator, translation stage, LDV and accelerometer. While several experiments may be 'executed' remotely once the equipment parameters have been optimized in person, in this work, each tunable parameter of all the hardware can be adjusted remotely as well, therefore, eliminating that extra requirement.  The setup described is also modular, and has been used to analyze the modal response of samples actuated through air-based acoustic signals as well as those induced mechanically.\\
In the setup described in this work, an arbitrary waveform is set in a function generator, which is then amplified and played through a speaker in an opened transmission loss tube.  The acoustic signal from the speaker travels through the air in the tube to vibrate the sample surface.  The LDV is mounted on a 2D motorized platform, which is controlled through LabVIEW on a desktop computer.  The LDV continuously sends velocity and signal strength data to a Compact Rio (cRio) data acquisition device, which records data upon trigger from the the function generator.  The recorded data is then streamed and saved on the desktop computer.  The Fourier transform of each measurement location is calculated in MATLAB, generating a 3D data array, consisting of two spatial dimensions and one dimension in frequency.  A single frequency slice of data is taken from this array, which gives the amplitude of a 2D surface at a single frequency, also called a mode shape.  These mode shapes can then be compared to expected results from COMSOL simulations.

\end{executivesummary}

\chapter{Introduction}
\section{Motivation}
The current COVID-19 pandemic has required adaptation to a rapidly changing situation and long-term telework.  To continue to do bench work, it makes sense to automate processes that are able to be automated, and allow for remote access and control of the testing environment.  For example, one of our experiments required finely spaced vibration measurements across the surface of a sample, with various input excitations, measurement sampling rates and spacing.  With only a single-point LDV system and a data acquisition system (DAQ), it would take many hours to take the measurements while also increasing the chance of human error.  Instead of the experiment operator manually changing the excitation and measurement parameters, it is more efficient and convenient to automate, remotely control and monitor the experiment.  Therefore, the function generator parameter setup was automated and integrated with the pre-pandemic point-scanning setup.

An LDV system is widely used to measure the response of a system/sample to an excitation. Such experiments involve three broad stages: (1) sample fabrication and mounting, (2) excitation and (3) measurement scans.  The sample preparation is entirely problem-specific and is usually not integrable with the latter two. A variety of different excitation mechanisms/inputs maybe employed for experiments. In our lab, we perform vibrational or acoustic excitations.  Typically, once the sample is fabricated and mounted to the excitation mechanism in a location accessible to the LDV for scanning, it still requires a significant (at-work) time commitment to synchronize and optimize the excitation and LDV settings before the experiment can begin.

Here, our remote operation of the LDV system integrates all the components of the excitation mechanism as well, allowing the user to remotely control excitation and LDV settings to optimize them and run the measurement scan.  This includes control of the function generator, thereby adjusting the output of a vibrational exciter or sound tube.

Commercial solutions to just the scanning stage of the problem exist for some specific applications, such as the 2D and 3D LDV systems sold by various companies.  However, higher dimensional systems can cost on the order of 10 times more than a single-point system for each added dimension of measurement capability.  For many lab setups, this may not be feasible, and in many situations do not allow for enough adaptability and control of the scan parameters.  The methods described here can be reproduced at lower cost and less administrative time used to purchase a 2D LDV while still producing high quality results, in addition to customization and integration with the testing environment.  Another crucial advantage of the current approach is that a vendor-built custom 2D scanning system is wedded to just the LDV measurement, and has no use during the machine's down-time. Whereas, as a consequence of the present work, the 2D motorized stage can be used modularly for a wide range of other experiments, and the automation/remote-control components carry forward to several other case scenarios. For example, the motorized stage and function generator can be placed in an acoustic (hydrophone) scan experiment and moved back to the LDV setup as needed.

\section{Background}
Historically, in the Acoustic Signal Processing and Systems Branch (Code 7160), we performed experiments using scanning-type measurements for underwater acoustic tests, using an automated system to move and acquire data from a hydrophone in 2D and 3D scans.  The code used to run the underwater tests was modified to run on different hardware and connect to different systems, but the idea behind the measurement remains the same.  By moving the data acquisition location around, the data collected represents the same data that would have been collected if using a large array of measurement devices.  Conversely, the excitation source can be moved around with a constant data acquisition location, which would emulate having an array of sources.

The data collected at each measurement point is usually then processed in MATLAB to extract binned time series data or frequency domain data.  These analyses can then be represented as a propagating wave or modal shape, respectively.  As additive manufacturing and featuring capabilities achieve finer resolutions, the demand for higher frequency scans has only been increasing. Moreover, an interest in studying non-linear acoustic or transient phenomena also requires large, dense collections of data points and high-frequency scans. These result in a substantial increase in the amount of data acquired per scan, necessitating parallel processing of the data processing step.

\section{A Brief Overview of This Work}
\subsection{Input Excitation}
The testing setup described in this work sets an arbitrary waveform in a function generator, which is then amplified and played through a speaker in an opened transmission loss tube.  The acoustic signal from the speaker travels through the air in the tube to vibrate the sample surface.  The transmission tube can be replaced by a vibration exciter, if the desired actuation is mechanical rather than pressure based.
\subsection{Output Measurement}
The LDV is mounted on a 2D motorized platform, which is controlled through LabVIEW on a desktop computer.  The LDV continuously sends velocity and signal strength data to a cRio data acquisition device, which records data upon trigger from the the function generator.  At the same time, the data from the accelerometer is also collected, so that the output can be normalized for each measurement location.  The recorded data is then streamed and saved on the desktop computer.  Since the controls for this setup are all on the desktop personal computer (PC), this allows remote desktop users to modify the input excitation, linear stages and DAQ parameters while teleworking.
\subsection{Data Analysis}
\begin{figure}
\includegraphics[width=\linewidth]{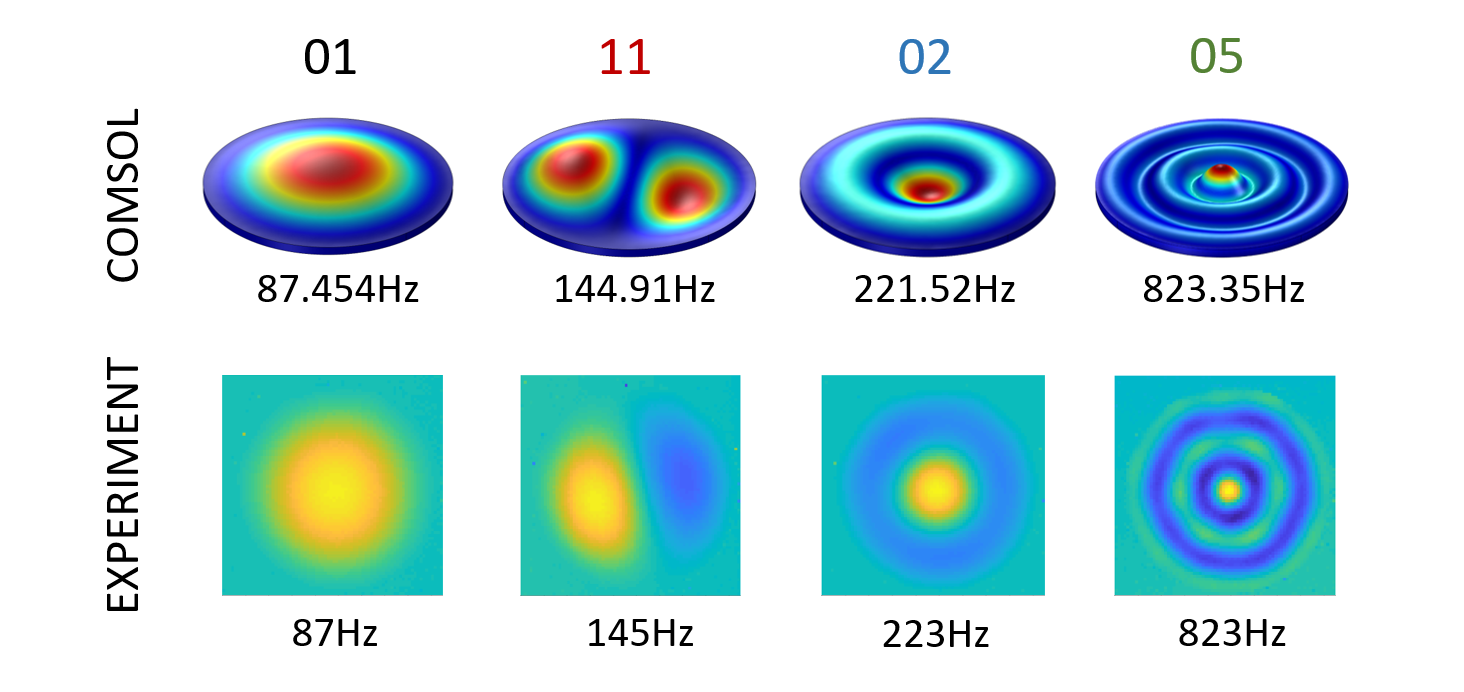}
\caption{Vibrational modes of a circular elastomeric plate, produced using the setup described in this work and compared with a COMSOL model.  Figure reused with permission \cite{wissman2019soft} \cite{wissman2019liquid}.}
\label{fig.modeshapes}
\end{figure}

The Fourier transform of each measurement location is calculated in MATLAB, generating a 3D data array.  A single frequency slice of data is taken from this array, which gives the amplitude of a 2D surface at a single frequency, also called a mode shape.  These mode shapes can then be further analyzed and compared to expected results in COMSOL simulations, as seen in Fig. \ref{fig.modeshapes}.

\pagebreak
\chapter{Equipment and Connections}
The desktop PC was connected through a Universal Serial Bus (USB) cable to the function generator.  The signal output of the function generator was connected to the input of the amplifier through a Bayonet Neill-Concelman (BNC) cable.  The output of the amplifier was connected to the input of the tube through banana cables.  The impedance/sound tube kit can be substituted with a vibrational exciter when mechanical actuation is required instead.

The PC was connected through RS-232 to the 2D stage controller.  The LDV was mounted on a custom mounting bracket to the stage.  The PC was connected to the DAQ through an Ethernet cable, and the DAQ modules were inserted into the DAQ chassis.  The signal strength and velocity outputs of the LDV were connected to the analog input of the DAQ.

\begin{figure}
\includegraphics[width=\linewidth]{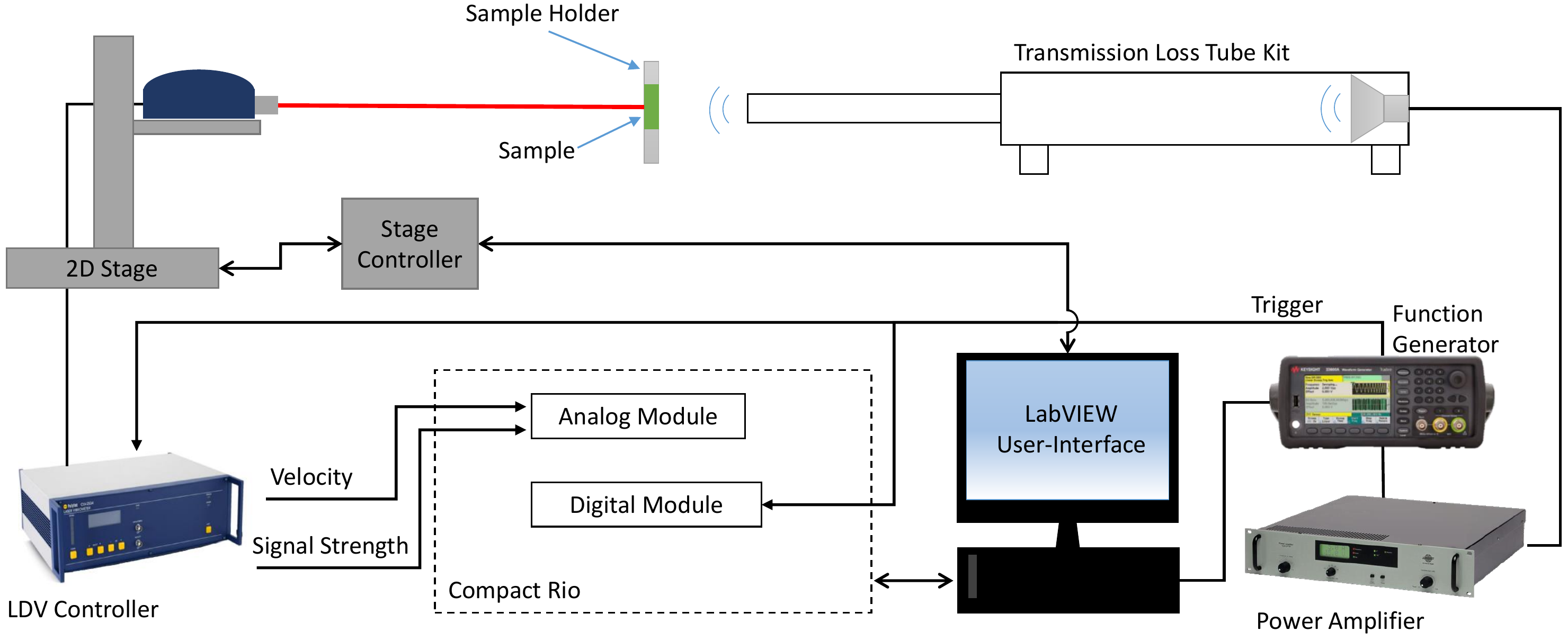}
\caption{Equipment connections used to perform 2D LDV scan of the elastomeric plate seen in Fig \ref{fig.modeshapes}.}
\label{fig.setup}
\end{figure}

The trigger signal from the function generator was connected through BNC to the digital module on the DAQ as well as the anti-drift input on the LDV controller.  This causes the voltage output of the LDV to be set to 0V at each trigger, which helps to avoid the signal exceeding the maximum allowed by the LDV controller (10V).  The trigger is also used by the DAQ to determine when to start recording the signal for each measurement location.  The equipment and connection schematics are illustrated in Fig. \ref{fig.setup}, and their make and model are listed in Table \ref{tab.equipment}.

\begin{center}
	\begin{table}
		\begin{tabular}{ |c|c|c| } 
			\hline
			\textbf{Equipment} & \textbf{Model} & \textbf{Manufacturer}\\ \hline
			Function Generator & 33500B & Agilent Technologies\\ \hline
			Power Amplifier  & Type 2718  & Br\"{u}el \& Kj\ae r (B\&K)\\ \hline
			Transmission Loss Tube Kit & Type 4206 & Br\"{u}el \& Kj\ae r (B\&K)\\ \hline
			2D Linear Stage & Bi-Slide & Velmex Inc.\\ \hline
			Single-Point LDV & CLV-2534 & Polytec Gmbh\\ \hline
			DAQ Chassis & cRio-9035 & National Instruments\\ \hline
			Analog Input Module & NI-9223 & National Instruments\\ \hline
			Digital I/O Module & NI-9402 & National Instruments\\ \hline 
		\end{tabular}
		\label{tab.equipment}
		\caption{List of equipment and their respective makes and models, used to take a 2D LDV scan of the elastomeric plate seen in Fig \ref{fig.modeshapes}.}
	\end{table}
\end{center}

\chapter{Data Acquisition Using LabVIEW}
The description here assumes basic familiarity with LabVIEW.  The code is run from several systems: the desktop PC, the Real-Time (RT) controller, and the FPGA inside of the controller chassis.  The LabView code uses the FPGA and the C-series DAQ modules on the cRio to take data, passes it along through the RT controller and then the PC.  In addition to the base LabVIEW program, the Embedded Control Suite is required to operate the FPGA.  A flow chart showing the communication logic is displayed in Fig. \ref{app.flowchart}.  LabVIEW block diagrams are shown in Appendix \ref{app.labview}.
\section{FPGA}
In the FPGA code, the digital input is constantly read until it detects a rising edge from the function generator trigger.  It then takes the analog voltage time series through a for-loop, and passes the values into a Direct Memory Access (DMA) first-in first-out (FIFO) channel.  DMA FIFO channels are a way to communicate data at high sampling rates between the FPGA and the RT controller.  The number of iterations of the for-loop and the timing between iterations is set by a control.  Control values are modifiable from the RT code.  The LDV signal strength is passed into an indicator.  The DMA FIFO is also checked if it overflowed, and this value is passed onto an indicator.  The overflow should be incorporated with a feedback node, so that successive reads of the overflow indicator do not reset the value.  The FPGA code can be seen in Figure \ref{app.fpga}.
\section{RT Controller}
The RT controller opens the FPGA program, and sets the FPGA control values using the ``Read/Write Control'' function.  The RT controller waits until the 2D stage has stopped, which is determined by a Boolean shared variable.  Before and after data acquisition, the RT controller code checks if the 2D stage has stopped moving, and tells the PC that it is done acquiring data for that sampling location through Boolean shared variables.  The FPGA code is then started, and the data is read from the DMA FIFO through a while loop.  The while loop queries the DMA FIFO to check if there are elements stored on it.  If there are, it reads the elements and places it in an array.  The array is appended to previous iterations of the while loop.  When the while loop is done, the array contains data from a single average.  The for loop that surrounds the while loop repeats the single average process until it contains the data from the number of averages requested, as seen in Fig. \ref{app.RT}.
\section{Desktop PC}
The desktop PC controls the 2D stage and the function generator, and coordinates with the RT controller to acquire data between movements of the LDV.  In the first section of the PC code, the 2D linear stage is initialized, which gives the stage controller an origin to reference.  The initialization sub-function of the LabVIEW Virtual Instrument (sub VI) is custom made, based on communication protocols supplied by the manufacturer.  The function generator parameters and the arbitrary waveform file are read into a custom made sub VI, which utilizes drivers from the manufacturer.  This is shown in Fig. \ref{app.pc-1}.

In the next selected snippet, the PC code creates a raster grid based on user input in Fig. \ref{app.pc-2}.  The raster grid is iterated through nested for loops, reading and saving data each time to the hard drive, as seen in Fig. \ref{app.pc-3}.  The PC tells the RT controller when the movement is done through a Boolean shared variable, and waits for the data acquisition to finish before reading the data.  Once it finishes saving, it moves on to the next iteration of the nested for loop.  When the nested for loop is done, it tells the cRio that the scan is done, and closes the communication port for the function generator.  Failure to close ports properly as mentioned here, will lead to communication issues in subsequent operation, which can usually be remedied by resetting the device.

Some of the default settings on the function generator are not suited for our application.  For example, the function generator applies a low pass filter at 17kHz, which limits the types of response that can be induced in the sample.  By including code that can modify parameters on the function generator, these settings can be turned off, modified and automated, allowing for different parameter sets to be used when acquiring multiple data sets.

\chapter{Data Analysis in MATLAB}
\section{Loading and Processing Data}
The data taken during the experiment was saved in a folder as separate comma delimited text files.  In MATLAB, the working directory was changed to the folder containing the data files.  Since the script described here used parallel processing to increase the processing speed, the MATLAB parallel processing toolbox is required to run it.  In the script, the file names in the directory are read into an array, and the X and Y values are extracted from their names into two 1D arrays.  The frequency array is calculated based on the sampling rate and the amount of zero padding that is desired.  Zero padding refers to a signal processing technique in which zeros are added to the original time series data, which decreases the discrete frequency step size in the corresponding fast Fourier Transform (FFT).  While increasing the zero padding of a data set provides an increasingly small frequency step size, the resulting FFT calculations taking increasingly more time to calculate, and therefore there is an inherent trade-off between the acceptable processing time and the minimum frequency step size achievable.  A parallelized for (parfor) loop is then used to read the data and perform an FFT on all the files in the directory, resulting in a 2D array (each column is the FFT of a different position, and each row represents a different frequency).

\section{Using FFT Data to Perform Modal Analysis}
In the next section of the script, a new directory is created, which is used later to save the figure files in.  The frequency limit of the mode shapes to be displayed are set here, because the frequency range of interest is usually less than what the entire FFT contains.  Using a parfor loop again, the images are generated from the FFT data.  The X and Y arrays are used to index the FFT array, and the iteration counter of the parfor loop is used to select the row, which generates a 2D array of amplitudes that represent a single frequency.  The 2D array is then plotted as a color map or a 3D mesh figure.  Using this code, 500 images containing 143 by 136 spatial points (representing 19,448 sampling locations) were created in about 2 minutes on a computer with a 16 core processor.  Further optimization can be carried out by doing parallel analysis on a Graphics Processing Unit (GPU) rather than on a Central Processing Unit (CPU) as was done here, which would allow many additional parallel processes, resulting in lower computation time.

In comparison, without a parallel processing approach the code would have to run on a single processor, which would likely take approximately 16 times as long, saving half an hour for a relatively small subset of the data.  With larger datasets or more detailed binning of the Fourier transform, the time saved would likely increase linearly, with a similar factor of performance improvement.  The MATLAB code is included in Appendix \ref{app.matlab}, and the custom functions used are included in Appendix \ref{app.fft} and \ref{app.velocity}.
\chapter{Discussion}
The method described here shows that a 2D LDV measurement can be done well with a single-point LDV and easily obtained components.  The modular nature of the setup allows it to be used for different applications; to actuate a thin elastomeric plate as seen here through pressure waves in air, or through mechanical means to vibrate more rigid samples, using a vibration exciter and a stinger.  In addition, some components can be swapped out for cheaper parts, especially the DAQ.  A simple FPGA based DAQ card can be obtained on the order of \$100, which is two orders of magnitude cheaper than a cRio.  Uniquely for NRL, and perhaps many other research laboratories, FPGA DAQs are usually present with a lot of down time.  The modularity of this solution can be extended to various existing hardware.

Conversely, to acquire data at faster speeds, better damping on the linear stages and increasing its rigidity can be done to decrease the time needed for each sampling location.  The number of averages and the number of time samples can also be modified to increase acquisition speed.  However, this can affect the quality of the data produced, since noise generally decreases with the square root of the number of averages taken.

Another important factor to consider is the sample mounting.  Poorly mounted samples can lead to poor actuation of the sample.  The sample shown in Fig. \ref{fig.modeshapes} has the elastomer bonded with an acrylic frame, which helps to ensure more uniform actuation.  Especially in mechanical actuation (e.g. actuation of more solid objects, like plastic or metal plates), the contact between the vibration exciter, the stinger, and the sample must be snug.

The signal strength value from the LDV system can also be used to filter out low signal sampling locations.  In many cases, signal strength varies arbitrarily due to surface roughness effects, and more often a region of good signal strength maybe found infinitesimally close to a spot with poor signal strength. Therefore, for these scenarios, the LabVIEW code could also be modified to step a small amount in the X or Y direction upon measurement of a poor signal location.  If these approaches are not feasible for a particular use case, then the data can also be spatially filtered to smooth out the figure surface.

\chapter{Conclusion}
To summarize the process described in this work, a sample is vibrated using an arbitrary waveform played through a speaker.  A 2D linear stage is used to move a single-point LDV, which takes the velocity time series of the sample's surface.  The time series is saved to individual text files, and the FFT is taken of each file.  For each frequency bin in the FFT, the sampling position is used to assign its position in a 2D mesh.  The mesh is then plotted either as a surface plot or as a colorplot, resulting in a mode shape plot.  By using parallel processing, this process is sped up significantly, resulting in large time savings which increase with the complexity and size of the analysis.  The modular nature of this setup also allows for modification and therefore versatility for different use cases and budgets.

In addition to the time savings and low cost, the automation and remote operation of its sub components makes this setup even more appealing in times where lab access is restricted.  As such, this report serves to document the technical process used to develop the measurement system to accelerate the development of similar setups and preserve institutional knowledge.

The extent of remote operation can be further improved from this setup by employing a third-axis translation stage (in the direction of the LDV scanning beam). For optimal LDV signal strength, it requires the sample to be separated at discretized distances from the beam output location. The third stage can help achieve this remotely, and eliminate the need for in-person adjustment of the sample and/or LDV.  Furthermore, when samples have a curvature, thereby varying that distance during a scan, the third axis translation stage can be used to dynamically adjust the LDV separation based on each sampling location to provide the best signal strength throughout the sample.

The high level of detail and discussion of the code used are included to help unfamiliar engineers to be brought up to speed quickly.  The LabVIEW and MATLAB codes used are shown in the appendices below, to give further guidance.

\begin{acknowledgments}
The LabVIEW code displayed here was built upon work performed by Dr. Michael Nicholas (NRL, Code 7165).  We thank Wissman et. al. for permission to reuse their figure.  We also thank Dr. Matthew Guild (NRL, Code 7165) and Dr. David Calvo (NRL, Code 7165) for their time spent reviewing this manuscript.
\end{acknowledgments}

\bibliography{2D_LDV_Report}

\begin{thebibliography}{2}
\providecommand{\natexlab}[1]{#1}
\providecommand{\url}[1]{\texttt{#1}}
\expandafter\ifx\csname urlstyle\endcsname\relax
  \providecommand{\doi}[1]{doi: #1}\else
  \providecommand{\doi}{doi: \begingroup \urlstyle{rm}\Url}\fi

\bibitem[Wissman et~al.(2019{\natexlab{a}})Wissman, Sampath, Ikei,
  {\"O}z{\"u}temiz, Majidi, and Rohde]{wissman2019soft}
J.~Wissman, K.~Sampath, A.~Ikei, K.\,B. {\"O}z{\"u}temiz, C.~Majidi, and C.\,A.
  Rohde,
\newblock ``Soft-matter pressure sensors for turbulence detection,''
\newblock Proceedings of the Sensors and Smart Structures Technologies for
  Civil, Mechanical, and Aerospace Systems 2019, volume 10970 (International
  Society for Optics and Photonics), 2019{\natexlab{a}}, p. 109702D.

\bibitem[Wissman et~al.(2019{\natexlab{b}})Wissman, Sampath, Ikei,
  {\"O}z{\"u}temiz, Carmel, and Rohde]{wissman2019liquid}
J.\,P. Wissman, K.~Sampath, A.~Ikei, K.\,B. {\"O}z{\"u}temiz, M.~Carmel, and
  C.~Rohde,
\newblock ``Liquid metal-based resistive membranes for flow acoustics
  detection,''
\newblock \emph{The Journal of the Acoustical Society of America} {\bf 146}(4),
  2997--2997 (2019{\natexlab{b}}).

\end{thebibliography}


\appendix

\chapter{LabVIEW Data Acquisition Code}
\label{app.labview}

\begin{figure}
\includegraphics[width=0.35\linewidth]{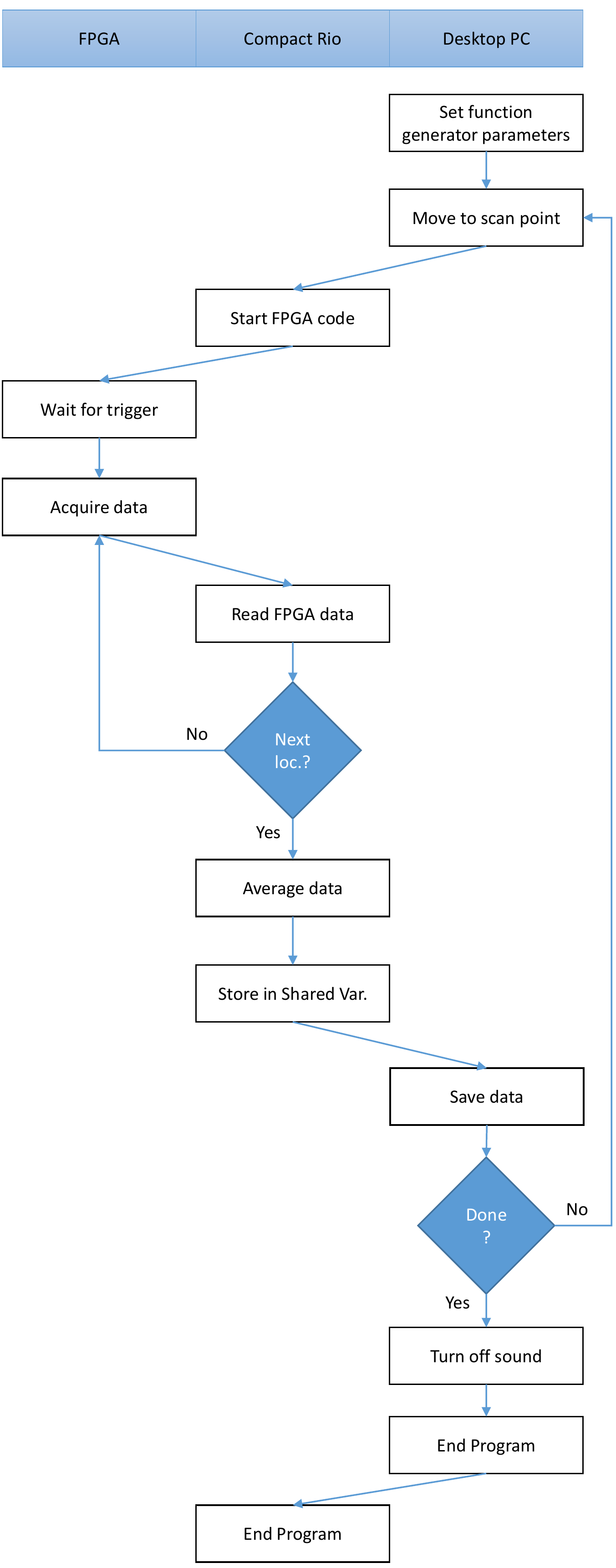}
\caption{Communication flow chart between systems running LabVIEW code.}
\label{app.flowchart}
\end{figure}

\begin{figure}
\includegraphics[width=0.5\linewidth]{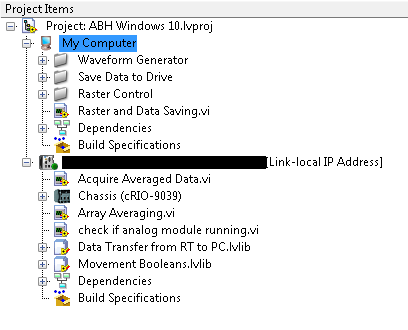}
\caption{Organization of LabVIEW files.  Opening the ".lvproj" file opens this LabVIEW project window.}
\end{figure}

\begin{figure}
\includegraphics[width=\linewidth]{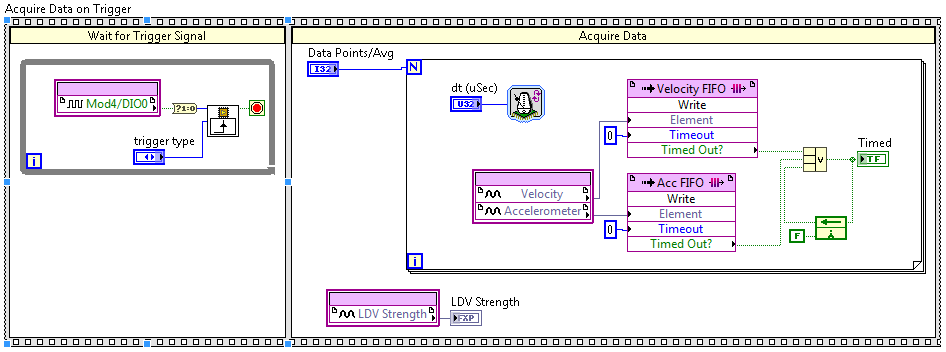}
\caption{FPGA code, located under cRio $\rightarrow$ Chassis $\rightarrow$ FPGA Target.  This code reads the digital input and moves to the second panel after receiving a rising edge.  The second panel contains a for loop that collects the analog voltage signal and passes the value into the DMA FIFO at specifically timed intervals.  If the FIFO overflows, it sets a boolean indicator to true.  The signal strength of the LDV is also recorded as a single value.  The controls can be adjusted from the RT controller code.}
\label{app.fpga}
\end{figure}

\begin{figure}
\includegraphics[width=\linewidth]{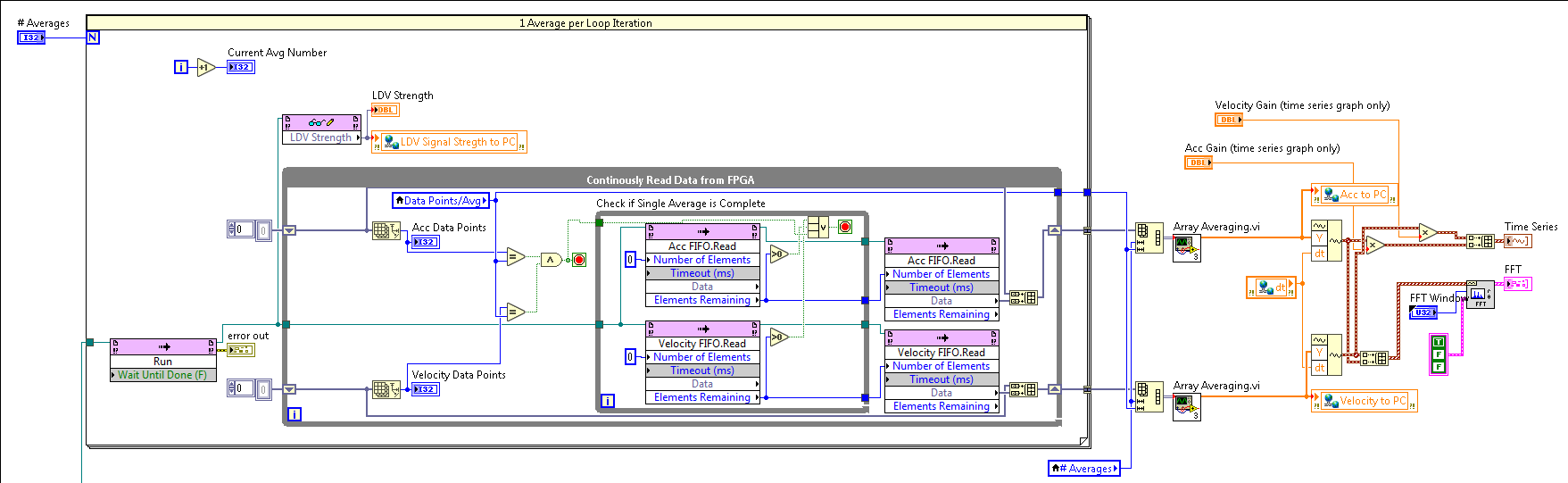}
\caption{Selected portion of the Real-Time Controller Code, located under cRIO.  This code runs the FPGA code, then reads the LDV strength into a shared variable.  The DMA FIFO for each channel is queried.  If there are elements remaining, they are read and passed out of the while loop into the for loop.  The for loop continues until the number of averages taken is complete.  The data is then averaged and passed into shared variables and the front panel display.}
\label{app.RT}
\end{figure}

\begin{figure}
\includegraphics[width=0.5\linewidth]{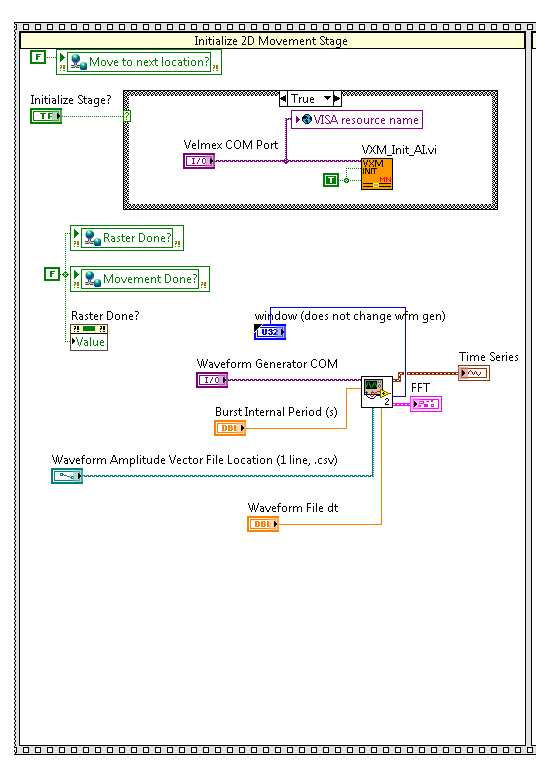}
\caption{First part of the PC code, located under My Computer.  This section initializes the 2D stage, so that it has an origin to reference.  The waveform generator is also loaded with the arbitrary waveform.}
\label{app.pc-1}
\end{figure}

\begin{figure}
\includegraphics[width=\linewidth]{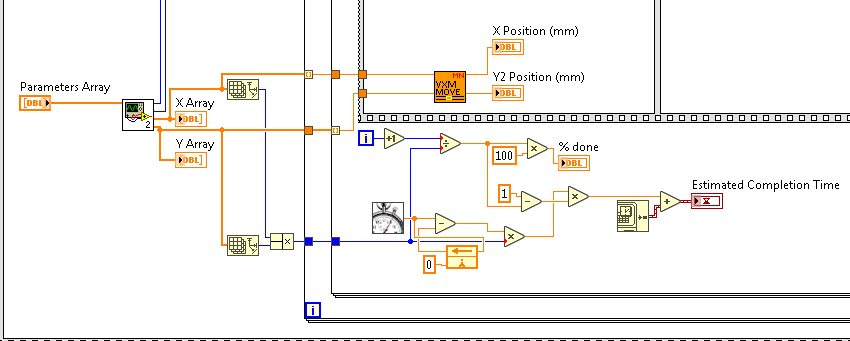}
\caption{Second part of the PC code, located under My Computer. This section generates the raster grid based on user input, and then moves the stage, iterating through nested for loops.}
\label{app.pc-2}
\end{figure}

\begin{figure}
\includegraphics[width=\linewidth]{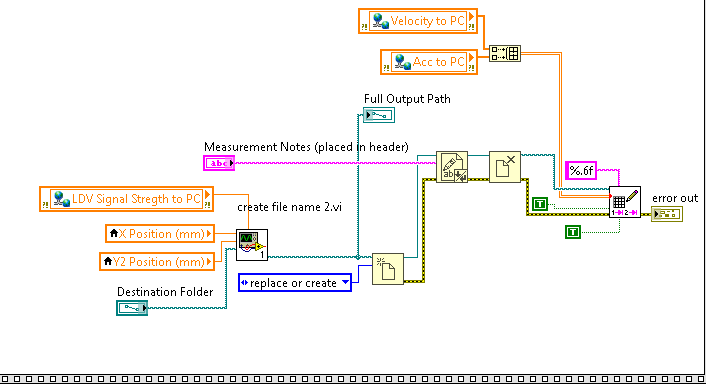}
\caption{Next important part of the PC code, located under My Computer. This section reads the data stored int he shared variables, and saves them to text files.  The precision is set by the format string "\%.6f".}
\label{app.pc-3}
\end{figure}

\chapter{MATLAB Parallel Processing 2D LDV Code}
\label{app.matlab}
\begin{verbatim}
%%Attribution
%Code written by Alec Ikei and Kaushik Sampath, November 2020
%Included in 2D LDV US Naval Research Laboratory Memorandum Report

%% Read Data and Parallel Calculate FFT
% get directory list
clc; clear; close all;
folder = uigetdir(); %prompts user to select folder containing data
fileList = dir(fullfile(folder, '*.tsv')); %only lists .tsv files
fileNames=strcat(folder,'\',{fileList.name}); %extract the names of the files and 
%put them into a n x 1 cell array

%User inputs
samplingRate=100000; %sampling rate used in Hz
padMultiplier=1; %number of TS lengths of 0s to pad
minFreq=100; maxFreq=20000; %set frequency limits for slicing
signalCutoff=0.05;
%end of user inputs

x=zeros(length(fileNames),1); y=x;z=x; badpxList=x; %length(a)=number of files
badpx=0; %# of low signal str pixels

%calculate the frequency vector for just one timeseries
velocityArray=importVelocity(fileNames{1}); %get LDV data for particular 
%spatial location
localfft=func_SSA(velocityArray,padMultiplier); %calculate FFT vector
numofBins=length(velocityArray)/2*padMultiplier; %calculate # frequency bins
localfreqBinSize=samplingRate*0.5/numofBins; %calculate bin width
freqArray=0:localfreqBinSize:samplingRate/2; %generate frequency array

%locate index of max frequency
    dist = abs(freqArray - maxFreq); 
    minDist = min(dist);
    maxFreqidx = find(dist == minDist);

%locate index of min frequency
    dist = abs(freqArray - minFreq); 
    minDist = min(dist);
    minFreqidx = find(dist == minDist); 

%Pre-allocate variable for FFT array
    fftArray=zeros(length(localfft(minFreqidx:maxFreqidx)),length(fileNames)); 
	%for FFT in columns

parfor m=1:length(fileNames) %parallel computation of FFTs
    m %shows current iteration in command window
    indexX=regexp(fileNames{m},'X_'); 
    indexY=regexp(fileNames{m},'Y2_'); 
    indexS=regexp(fileNames{m},'Strength'); 
    indexP=regexp(fileNames{m},'.tsv');%location from filename
    if str2double(fileNames{m}(indexS+8:indexP-1))>signalCutoff 
	% only use data if signal strength is good
        x(m)=str2double(fileNames{m}(indexX+2:indexY-1)); 
		% X coordinate from filename
        y(m)=str2double(fileNames{m}(indexY+3:indexS-1)); 
		% Y2 coordinate
        velocityArray=importVelocity(fileNames{m}); 
		%LDV data for particular spatial location
        localfft=func_SSA(velocityArray,padMultiplier); 
		%calculate FFT
        fftArray(:,m)=localfft(minFreqidx:maxFreqidx); 
		%place relevant FFT data in columns
    else
        badpx=badpx+1; %count number of excluded pixels
        badpxList(m)=1; %list bad pixels
    end
end

x(badpxList>0)=[];y(badpxList>0)=[];fftArray(:,badpxList>0)=[]; %remove bad pixels
fftArrayMax=max(max(abs(fftArray))); %global maximum amplitude
     
disp('done reading data');
%% Create color plot of each slice and save as .png
pngfolder=strcat(folder,'\pngs');figfolder=strcat(folder,'\figs'); 
%folders for images
mkdir(pngfolder); mkdir(figfolder);

xx=unique(x,'rows');yy=unique(y,'rows'); %remove duplicate coordinates
%creates a 2d set of points to assign the sampNormPlt values to
[X,Y]=meshgrid(min(xx):xx(2)-xx(1):max(xx),min(yy):yy(2)-yy(1):max(yy)); 

%pre-allocate space to save all surface data
surfaceData=zeros(length(minFreqidx:maxFreqidx),length(yy),length(xx)); 

currentFreq=minFreqidx:maxFreqidx;
parfor k=1:length(currentFreq) %must start at 1, since parfor
    f=scatteredInterpolant(x,y,fftArray(currentFreq(k),:)'); 
	%create function from data
    Z=f(X,Y)-mean(f(X,Y)); %create surface values from function, minus DC offset
    surf(X,Y,real(Z),'LineStyle','none'); %create surface plot
    view(2); %flat view
    imageName=strcat(num2str(freqArray(currentFreq(k))), ' Hz');
    print('-dpng','-r75',strcat(pngfolder,'\',imageName)); 
	%convert current fig to png
    saveas(gcf, strcat(figfolder,'\',imageName)); 
	%saves the current figure as Matlab file
end
\end{verbatim}

\chapter{MATLAB Custom Function: func\_SSA}
\label{app.fft}
This function takes the single sided amplitude complex Fourier transform.  It also can zero pad the timeseries input, to change the frequency binning.  Used in the parallel processing 2D LDV code.
\begin{verbatim}
function [SSA,Phase] =  func_SSA(X,k)
%k=1; changed from kaushiks code
% Sampling Length
L = length(X);
% Pad with a length of k times L
Lpad = k*L;X1 = zeros(Lpad,1);
X1(Lpad/2-L/2+1:Lpad/2+L/2)=X;
X=X1;

% Compute FFT
FFT1 = fft(X);
Phase=imag(FFT1);
% Single Sided Amplitude Spectrum
FFT2 = FFT1/Lpad;
SSA = FFT2(1:Lpad/2+1,:);
SSA(2:end-1,:) = 2*SSA(2:end-1,:);
SSA = SSA/(L/Lpad);
\end{verbatim}

\chapter{MATLAB Custom Function: importVelocity}
\label{app.velocity}
This function reads the text data file and converts it into a vector.  Used in the parallel processing 2D LDV code.
\begin{verbatim}function velocityArray = importVelocity(filename, dataLines)
if nargin < 2
    dataLines = [1, Inf];
end

%% Setup the Import Options
opts = delimitedTextImportOptions("NumVariables", 2);

% Specify range and delimiter
opts.DataLines = dataLines;
opts.Delimiter = "\t";

% Specify column names and types
opts.VariableNames = ["VarName1", "Var2"];
opts.SelectedVariableNames = "VarName1";
opts.VariableTypes = ["double", "string"];
opts = setvaropts(opts, 2, "WhitespaceRule", "preserve");
opts = setvaropts(opts, 2, "EmptyFieldRule", "auto");
opts.ExtraColumnsRule = "ignore";
opts.EmptyLineRule = "read";

% Import the data
tbl = readtable(filename, opts);

%% Convert to output type
velocityArray = tbl.VarName1;
end
\end{verbatim}
\end{document}